\documentclass[bm]{epl}
\usepackage{amssymb,epsfig}
\begin{document}

\title{Fingerprints of classical diffusion in open $2D$ mesoscopic systems in the metallic regime}
\shorttitle{Quantum scattering}

\author{A. Ossipov, Tsampikos Kottos, and T. Geisel
}
\institute{
\mbox{Max-Planck-Institut f\"ur Str\"omungsforschung und Fakult\"at Physik
der Universit\"at G\"ottingen,}\\
\mbox{Bunsenstra\ss e 10, D-37073 G\"ottingen, Germany}
}
\pacs{03.65.Nk}{Scattering theory}
\pacs{05.45.Mt}{Quantum chaos; semiclassical methods}
\pacs{73.23.-b}{Electronic transport in mesoscopic systems}

\maketitle

\begin{abstract}
We investigate the distribution of the resonance widths ${\cal P}(\Gamma)$ and Wigner 
delay times ${\cal P}(\tau_W)$ for scattering from two-dimensional systems in the 
diffusive regime. We obtain the forms of these distributions (log-normal for large
$\tau_W$ and small $\Gamma$, and power law in the opposite case) for different symmetry 
classes and show that they are determined by the underlying diffusive classical dynamics. 
Our theoretical arguments are supported by extensive numerical calculations.
\end{abstract}


Quantum scattering has been a subject of intensive research activity both 
in mesoscopic physics and in Quantum Chaos in the last 
years~\cite{S89,FS97,BGS91,TF00,TC99,SOKG00,KW02}. Among the most interesting quantities 
for the description of a 
scattering process are the Wigner delay times and resonance widths. The former quantity 
captures the time-dependent aspects of quantum scattering. It can be interpreted as 
the typical time an almost monochromatic wave packet remains in the interaction region. 
It is related to the energy derivative of the total phase shift $\Phi(E)=-i\ln \det 
S(E)$ of the scattering matrix $S (E)$, i.e.  $\tau_W (E) = {d\Phi (E) \over dE}$. 
Resonances are defined as poles of the $S$-matrix occurring at complex energies ${\cal 
E}_n = E_n - \frac i2 \Gamma_n$, where $E_n$ is the position and $\Gamma_n$ the width 
of the resonance. They correspond to "eigenstates" of the open system that decay in 
time due to the coupling to the "outside world".

For chaotic/ballistic systems Random Matrix Theory (RMT) is applicable, and the 
distributions of resonance widths ${\cal P}(\Gamma)$ and Wigner delay times ${\cal P}
(\tau_W)$ are known \cite{FS97}. As the disorder increases the system becomes diffusive 
and the deviations from RMT become increasingly apparent. In the strongly disordered 
limit where localization dominates, the distribution of resonances ${\cal P} (\Gamma)$ 
\cite{TF00} and delay times ${\cal P}(\tau_W)$ \cite{TC99} were found recently. At the 
same time, an attempt to understand systems at critical conditions, was undertaken in 
\cite{SOKG00,KW02}. For diffusive mesoscopic samples, however, there is no study of 
${\cal P}(\Gamma)$ and ${\cal P}(\tau_W)$ besides Ref.~\cite{BGS91} where the authors 
have focused on the tails of ${\cal P}(\Gamma)$ for a quasi-1D system in the diffusive 
regime. This study is important for diffusive random lasers, where the knowledge of 
short resonance width distribution determines the properties of lasing thresholds 
\cite{WAL95}, as well as for various other applications like mesoscopic capacitors 
\cite{GMB96}, microwave cavities \cite{S99} and chaotic optical cavities \cite{NS97} 
where most of the theoretical treatment is limited by RMT. 
We point here that current developments of microwave experiments in random dielectric 
media in the diffusive regime \cite{CZG03} may allow us a direct comparison between
theory and experiment. Finally we note that an analogous theoretical study for the 
distribution of wavefunction intensities of {\it closed} systems was put forward very 
early in the development of the mesoscopic physics \cite{AKL91} and recently their 
statistical properties have become clear \cite{MK95,M97,N01,OKG02,ARS01}. 

In this paper, we study ${\cal P}(\Gamma)$ and ${\cal P}(\tau_W)$ for $2D$ open systems 
in the diffusive regime. We will show that these distributions are determined by the 
diffusive classical dynamics of the corresponding closed system and depend on the 
time-reversal symmetry (TRS). Specifically, the resonance width distribution ${\cal P}
(\Gamma)$ is given by
\begin{eqnarray}
\label{gammas}
{\cal P}(\Gamma < \Gamma_{cl}) & \sim & \exp(-C_\beta (\ln\Gamma)^2)\,, 
\,\,\,\,\,{\rm where}\,\,\,\,\,
C_{\beta}\sim \beta D \nonumber\\
{\cal P}(\Gamma \gtrsim \Gamma_{cl}) & \sim & {\sqrt {D\over L^2}}{1\over \Gamma^{3/2}}
\,\,\,\,\,\,\,\,
\end{eqnarray}
where $D$ is the classical diffusion coefficient (which is proportional to the Thouless 
conductance for disordered $2D$ systems), $L$ is the linear size of the system, $\Gamma_{cl}$ 
is the classical decay rate \cite{note0} which is inversely proportional to the Thouless 
time $L^2/D$ and $\beta$ denotes the symmetry class [$\beta=1 (2)$ for preserved (broken) 
TRS]. For the distribution of Wigner delay times we obtain
\begin{eqnarray}
\label{taus}
{\cal P}(\tau_W\lesssim \Gamma_{cl}^{-1}) & \sim & {1\over\tau_W^{3/2}}\exp(-\sigma/\tau_W)\,\,\,\,\,\,\,
\,\,\,\, \,\, \nonumber\\
{\cal P}(\tau_W > \Gamma_{cl}^{-1}) & \sim & \exp(-C_\beta (\ln\tau_W)^2)\,\,\,
\end{eqnarray}
where $\sigma$ is some constant of order unity. Our results (\ref{gammas},\ref{taus}) 
are applicable both for chaotic and disordered systems, provided that: (i) transport 
between scattering events may be treated semiclassically, (ii) a particle scatters many 
times while it traverses the system and (iii) strong localization effects can be neglected.

To numerically demonstrate the above predictions we use a model of the $2D$ Kicked Rotor 
(KR) with absorbing boundary conditions. The corresponding closed system is described 
by the time-dependent Hamiltonian \cite{OKG02,DF88}
\begin{eqnarray}
\label{ham}
H=H_0 + k V\sum_m \delta (t-m)\,\,, H_0(\{{\cal L}_i\}) = \sum_{i=1}^2
\frac{\rho_i}{2} {\cal L}_i^2\, \quad \quad \quad \quad \quad \quad
\quad \quad \quad \quad\quad \quad \quad \quad \\
V(\{\theta_i\}) = \cos(\theta_1) \cos(\theta_2) \cos({\alpha}) +
{1\over2} \sin(2\theta_1) \cos(2\theta_2) \sin({\alpha}) \quad \quad
\quad \quad \quad \quad\quad\quad \quad \quad \quad \quad \quad \nonumber
\end{eqnarray}
where ${\cal L}_i$ denotes the angular momentum and $\theta_i$ the conjugate angle of 
one rotor. Here the kick period is one, $k$ is the kicking strength, while $\rho_i$ 
is a constant inversely proportional to the moment of inertia of the rotor. The parameter 
$\alpha$ breaks TRS. The motion generated by (\ref{ham}) is classically chaotic and for 
a sufficiently strong kicking strength $k$ there is diffusion in momentum space with 
diffusion coefficient $D\equiv lim_{t\rightarrow \infty} <{\bf {\cal L}}^2(t)>/ t 
\simeq {k^2\over 2}$ \cite{DF88}.

Quantum mechanically, this system can be described by a finite-dimensional evolution 
operator for one period
\begin{equation}
\label{Uop}
U=e^{(-iH_0(\{{\cal L}_i\})/2)}\,\,e^{(-iV(\{\theta_i\}))} \,\,e^{(-iH_0(\{{\cal L}_i\})/2)}\,\,\,\,
\end{equation} 
where we put $\hbar=1$. The eigenvalues of $U$ are $\lambda_n=e^{i\omega_n}$ and
lie on the unit circle; $\omega_n$ are known as quasi-energies and are dimensionless.
The corresponding mean quasi-energy spacing is $\Delta = 2\pi/L^2$, where $L$ is the 
linear size of the system. A detailed presentation of our model can be found in 
\cite{OKG02}, where the statistical properties of wavefunction intensities in the 
regime of interest were investigated and found to be in agreement with the theoretical 
predictions for mesoscopic diffusive samples \cite{MK95,M97}. Therefore the $2D$ KR 
is a representative model both for $2D$ disordered and chaotic systems.

In this paper, we turn the closed $2D$ KR model into an open one. To this end we
impose absorption at the boundary of a square sample of size $L\times L$ in the 
momentum space. In other words, every time that one of the components of the two
dimensional momentum $({\cal L}_1,{\cal L}_2)$ takes on the value $1$ or $L$, the 
particle is absorbed without coming back to the sample. Using a recently proposed 
recipe \cite{FS00} we can write down the corresponding scattering matrix $S$ in 
the form \cite{O02}
\begin{equation}
\label{smatrix}
S(\omega) =  -W U e^{i\omega} {1\over I-e^{i\omega} P U} W^{\dagger} \,\,, 
P=I-W^{\dagger}W
\end{equation}
where $I$ is the $L^2\times L^2$ unit matrix and $W$ is a $M \times L^2$ matrix. It 
has only $M$ non-zero elements which are equal to one and describe at which "site" 
of the $L\times L$ sample we attach $M$ "leads" [in our case $M=4(L-1)]$. Here $W^{
\dagger}W$ is a projection operator onto the boundary, while $P$ is the complementary 
projection operator. The scattering matrix $S_{ij}$ given by Eq.~(\ref{smatrix}) can 
be interpreted in the following way: once a wave enters the sample, it undergoes 
multiple scattering induced by $[I-e^{i\omega} P U]^{-1}=\sum_{n=0}^{\infty} 
\left( e^{i\omega} PU\right)^n$ until it is transmitted out. It is clear therefore 
that the matrix ${\tilde U}=P U$ propagates the wave inside the sample. However, 
contrary to the closed system in which the evolution operator is unitary, the 
absorption breaks the unitarity of the evolution matrix ${\tilde U}$ so that all 
eigenvalues ${\tilde \lambda}$ move inside the unit circle. Therefore each eigenvalue 
can be written in the form ${\tilde \lambda}_n = e^{i{\tilde \omega_n}}= \exp(-i
\omega_n-\Gamma_n/2)$ where $\Gamma_n>0$ is the dimensionless resonance width of 
an eigenstate.

The Wigner delay time can be expressed as the sum of proper delay times $\tau_q$. The
latter are the eigenvalues of the Wigner-Smith operator written in our case as \cite{O02}
\begin{equation}
\label{qmatrix}
Q(\omega)\equiv {1\over i} S^{\dagger}{d S\over d \omega} = -e^{-i\omega} W 
K^{\dagger} W^{\dagger} W K U^{\dagger} K W^{\dagger}
\end{equation}
where $K=\left(P-U^{\dagger}e^{-i\omega}\right)^{-1}$. 

Below we present our theoretical considerations and compare them with the numerical
data obtained for the $2D$ KR model. The parameters of the model were chosen in such 
a way that the conditions (i)-(iii) discussed above were fulfilled (see 
\cite{OKG02}). In order to improve our statistics, we randomized the phases of the 
kinetic term of the evolution operator (\ref{Uop}) and used a number of different 
realizations. In all cases we had at least 60000 data for statistical processing. 

\begin{figure}
\centerline{\epsfig{figure=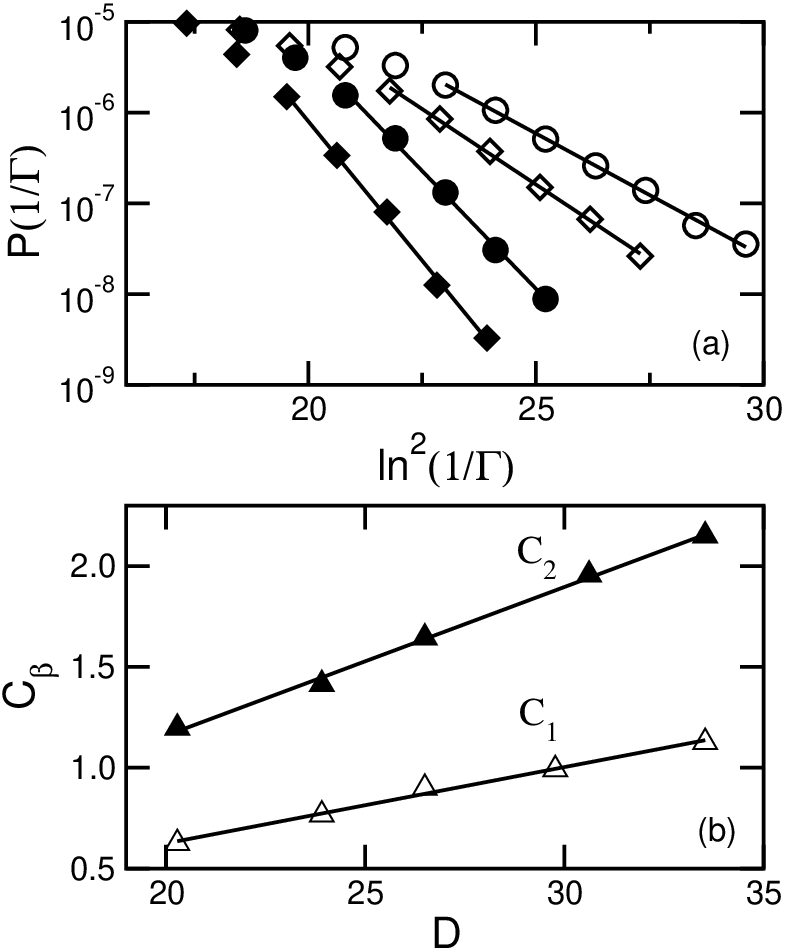,scale=0.8}}
{\footnotesize  FIG.1:
(a) The distribution of resonance widths (plotted as ${\cal P}(1/\Gamma)$
vs. $1/\Gamma$) for $\Gamma < \Gamma_{cl}$ for two representative values of $D$. The 
system size in all cases is $L=80$. Filled symbols correspond to broken TRS. The solid lines 
are the best fit of Eq.~(\ref{gammas}) for $\beta = 1 (2)$ to the numerical data. (b) 
Coefficients $C_{\beta}$ vs. $D$. The solid lines are the best fits to $C_{\beta}= 
A_{\beta} D+ B_{\beta}$ for $\beta=1 (2)$. The ratio $R=A_2/A_1 = 1.95\pm 0.03$}
\end{figure}

{\it Resonance widths distribution.} We start our analysis with the study of resonance 
width distribution ${\cal P}(\Gamma)$ for $\Gamma < \Gamma_{cl}$. 
The small resonances $\Gamma < \Delta$ can be associated, with the existence of 
pre-localized states of the "isolated" system. The latter states show strong localization 
features in contrast to the typical states in the diffusive regime and are considered 
as precursors of localization. They consist of a short-scale bump (where most of 
the norm is concentrated) and they decay rapidly in a power law from 
their center of localization \cite{MK95,M97}. One then expects that states of this 
type with localization centers at the bulk of the sample are affected very weakly by
the opening of the system at the boundaries. In first order perturbation theory,
considering the opening as a small perturbation we obtain
\begin{equation}
\label{pertgamma}
{\Gamma\over 2} = \langle \Psi|W^{\dagger}W|\Psi\rangle =\sum_{n\in {\rm boundary}}|\Psi(n)|^2 \sim 
L |\Psi(L)|^2 
\end{equation}
where $|\Psi (L)|^2$ is the wavefunction intensity of a pre-localized state at the 
boundary. At the same time the distribution of $\theta=1/{\sqrt L}\Psi (L)$ for large 
values of the argument is found to be of log-normal type i.e. ${\cal P}(\theta) \sim 
\exp\left(-\pi^2Dln^2 (\theta^2)\right)$ \cite{M97}. Using this together with Eq.~(
\ref{pertgamma}) we get ${\cal P}(1/\Gamma) \sim \exp\left(-\pi^2Dln^2 (1/\Gamma)
\right)$. We would like to stress that the expression for ${\cal P}(\theta)$, must
be corrected by including the TRS factor $\beta$ in the exponent. This is due to the
fact that the Optimal Fluctuation Method, which was used to derive the above expression
for ${\cal P}(\theta)$, does not describe the effect of breaking TRS in a correct way
\cite{OKG02,ARS01}. Taking all the above into account we end up with the expression 
given in Eq.~(\ref{gammas}).  

The numerical data reported in Fig.~1 support the validity of the above considerations. 
However, we mention that the perturbative argument is valid only for the case of very 
small resonances i.e. $\Gamma < \Delta$, whereas our numerical data indicate that one 
can extend the log-normal behaviour of ${\cal P}(\Gamma)$ up to resonances with $\Delta 
< \Gamma < \Gamma_{cl} $. 

\begin{figure}
\centerline{\epsfig{figure=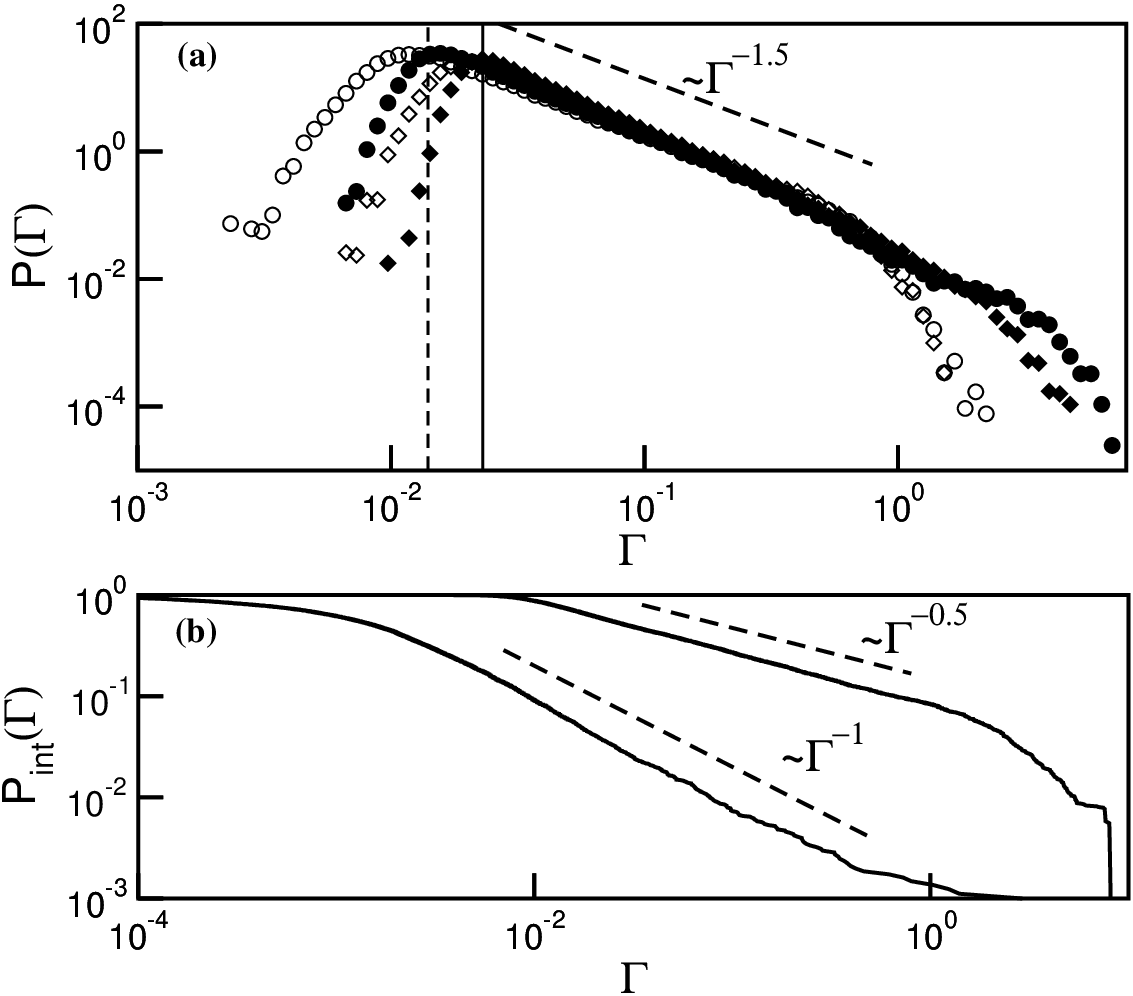,scale=0.8}}
{\footnotesize 
 FIG. 2: (a) Resonance width distribution ${\cal P}(\Gamma)$ for preserved
TRS and $D=20.3$ ($\circ$) and $D=33.5$ ($\Diamond$). The corresponding filled symbols
represent ${\cal P}(\Gamma)$ for broken TRS and the same values of $D$. The dashed 
(solid) vertical line mark the classical decay rate $\Gamma_{cl}$ \cite{note0} for 
$D=20.3 (33.5)$. (b) The ${\cal P}_{\rm int} (\Gamma)$ for a sample with nine leads 
(lower curve). For comparison we plot also the ${\cal P}_{\rm int}(\Gamma)$ for the same 
sample but when we open the system from the boundaries. The dashed lines correspond 
to the theoretical predictions (\ref{obound}) and (\ref{final}).}
\end{figure}

Next we turn to the analysis of ${\cal P}(\Gamma)$ for $\Gamma \gtrsim \Gamma_{cl}$. 
In this case, the resonances are overlapping, the corresponding eigenstates are strongly 
non-orthogonal and first-order perturbation theory breaks down \cite{FM02}. In Fig.~2a 
we report our numerical results for ${\cal P}(\Gamma)$ with preserved (broken) TRS for 
two representative values of $D$. An inverse power law ${\cal P}(\Gamma) \sim 
{\Gamma}^{-1.5}$ is evident in accordance with Eq.~(\ref{gammas}) \cite{note}. From 
Fig.~2a it is clear that this part of the distribution is independent of the symmetry 
class, in contrast to the small resonance distribution. 

The following argument provides some understanding of the behaviour of ${\cal P}(
\Gamma)$ for $\Gamma\gtrsim \Gamma_{cl}$. First we need to recall that the inverse of 
$\Gamma$ represents the quantum lifetime of a particle in the corresponding resonant 
state escaping into the leads. Moreover we assume that the particles are 
uniformly distributed inside the sample and diffuse until they reach the boundaries, 
where they are absorbed. Then we can associate the corresponding lifetimes with the
time $t_R\sim 1/ \Gamma_R \sim R^2/D$ a particle needs to reach the boundaries, when 
starting distance $R$ away. This classical picture can be justified for all states 
with $\Gamma \gtrsim \Gamma_{cl} \sim D/L^2$. The relative number of states that require 
a time $t<t_R$ in order to reach the boundaries (or equivalently the number of states 
with $\Gamma>\Gamma_R$) is 
\begin{equation}
\label{igam}
{\cal P}_{\rm int}(\Gamma_R)=\int_{\Gamma_R}^{\infty}{\cal P}(\Gamma)d\Gamma
\sim\frac{S(t_R)}{L^2}
\end{equation}
where $S(t_R)$ is the area populated by all particles with lifetimes $t<t_R$. In the 
case of open boundaries we get
\begin{equation}
\label{obound}
{\cal P}_{\rm int}(\Gamma_R)\sim \frac{L^2-(L-2R)^2} {L^2}\sim
\sqrt{\frac{\Gamma_{cl}}{\Gamma_R}}-\frac{\Gamma_{cl}}{\Gamma_R}.
\end{equation}
For $\Gamma_R > \Gamma_{cl}$ the first term in the above equation is the dominant one
and thus Eq.~(\ref{gammas}) follows.

Here it is interesting to point that a different way of opening the system might lead 
to a different power law behaviour for ${\cal P}(\Gamma)$. Such a situation can be 
realized if instead of opening the system at the boundaries we introduce "one-site" 
absorber (or one "lead") somewhere in the sample. In such a case we have 
\begin{equation}
\label{final}
{\cal P}_{\rm int}(\Gamma_R)\sim\frac{S(t_R)}{L^2}=\frac{R^2}{L^2}=\frac{Dt_R}{L^2}\sim
\frac{\Gamma_{cl}}{\Gamma_R}.
\end{equation}
The above result is valid for any number $M$ of "leads" such that the ratio $M/L^2$ scales
as $1/L^2$. In Fig.~2b we report ${\cal P}_{\rm int}(\Gamma)$ for the case with nine "leads" 
attached somewhere to the $2D$ sample. 

A straightforward generalization of our arguments for 3D systems in the metallic regime 
gives ${\cal P}_{\rm int}(\Gamma_R) \sim \left(\Gamma_{cl}/\Gamma_R\right)^{0.5} - 2 
\left(\Gamma_{cl}/\Gamma_R\right) +(4/3) \left(\Gamma_{cl}/\Gamma_R\right)^{1.5}$ which 
for $\Gamma_R > \Gamma_{cl}$ leads to the same universal expression as in Eq.~(\ref{gammas}). 
Similarly, the analogue of Eq.~(\ref{final}) in 3D is ${\cal P}_{\rm int}(\Gamma_R)\sim 
\left(\Gamma_{cl}/\Gamma_R\right)^{1.5}$.

{\it Wigner delay times distribution.} Our theoretical understanding will be based 
on the following relation 
\begin{equation}
\label{dtime}
\tau_W(\omega)=\sum_{n=1}^{L^2} \frac{\Gamma_n}{(\omega-\omega_n)^2 + \Gamma_n^2/4} 
\end{equation}
which connects the Wigner delay time and the poles of the $S-$matrix. Let us start with 
the far tails. It is evident that large times $\tau_W(\omega)\sim \Gamma^{-1}_n$ 
correspond to the cases when $\omega\simeq \omega_n$ and $\Gamma_n \ll 1$. Then for the 
distribution of delay times we obtain ${\cal P} (\tau_W) \sim \int d\Gamma {\cal P} 
(\Gamma)\Gamma \delta(\tau_W-1/\Gamma)$. Using the small resonance width asymptotics 
given by Eq.~(\ref{gammas}) we get the log-normal law of Eq.~(\ref{taus}). 

Now we estimate the behaviour of ${\cal P}(\tau_W)$ for $\tau_W \lesssim \Gamma_{cl}^{-1}$. 
In this regime many short-living resonances contribute to the sum (\ref{dtime}). We may
therefore consider $\tau_W$ as a sum of many independent positive random variables 
each of the type $\tau_n=\Gamma_n x_n$, where $x_n=\delta \omega_n^{-2}$. Assuming further
that $\delta \omega_n$ are uniformly distributed random numbers we find that the distribution
${\cal P}(x_n)$ has the asymptotic power law behaviour $1/x_n^{3/2}$. As a next step we 
find that the distribution ${\cal P}(\tau_n)$ decays asymptotically as $1/\tau_n^{3/2}$ 
where we use that ${\cal P}(\Gamma_n)\sim 1/\Gamma_n^{3/2}$. Then the corresponding ${\cal P}
(\tau_W)$ is known to be a stable asymmetric Levy distribution $L_{\mu,1}(\tau_W)$ of 
index $\mu=1/2$ \cite{BG90} which has the form given in Eq.~(\ref{taus}) at the origin. 
We point out here that the asymptotic behaviour ${\cal P}(\tau_W)\sim 1/\tau_W^{3/2}$ 
emerges also for chaotic/
ballistic systems where the assumption of uniformly distributed $\delta \omega_n$ is the 
only crucial ingredient (see for example \cite{FS97}). 

\begin{figure}
\centerline{\epsfig{figure=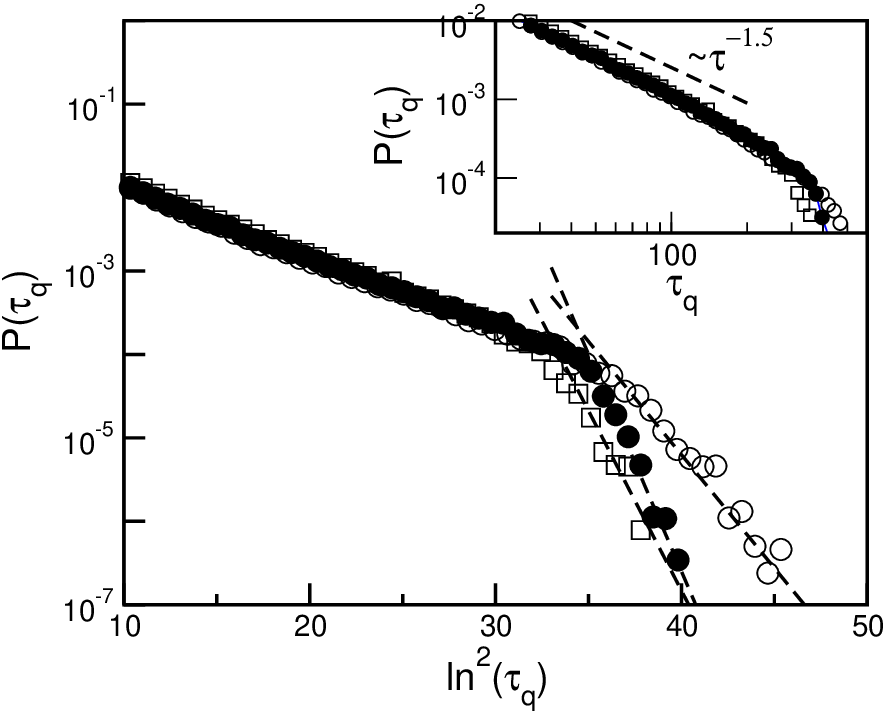,scale=0.8}}
{\footnotesize 
FIG. 3: The proper delay times distribution ${\cal P}(\tau_q)$ for $D=20.3 
(\circ)$ and $D=29.8 (\Box )$. The ($\bullet$) correspond to $D=20.3$ but now with 
broken TRS. The dashed lines have slopes equal to $C_{\beta}$ extracted from the 
corresponding ${\cal P}(\Gamma)$ (see Fig.~1b). In the inset we report ${\cal P}
(\tau_q)$ for moderate values of $\tau_q$ in a double logarithmic scale. }
\end{figure}

Since $\tau_W=\sum_{i=1}^M\tau_q$, we expect the behaviour of the distribution of proper 
delay times ${\cal P}(\tau_q)$ to be similar to ${\cal P}(\tau_W)$ for large values of 
the arguments (for $\tau_W\gg1$ we have $\tau_W\sim \tau_q^{\rm max}$). Moreover, from 
the numerical point of view ${\cal P}(\tau_q)$
can be studied in a better way because a larger set of data can be generated easily. 
Our numerical findings for ${\cal P}(\tau_q)$ are reported in Fig.~3 and are in nice
agreement with Eq.~(\ref{taus}), even for moderate values of $\tau_q$. We stress here
that the dashed lines in Fig.~3, have slopes equal to $C_{\beta}$ taken from the 
corresponding log-normal tails of ${\cal P}(\Gamma)$.

We thank R. Fleischmann and Y. Fyodorov for many useful discussions and comments and
C. Beenakker for pointing a misprint in Eq.~(\ref{Uop}). (T.K) thanks U. Smilansky for 
stimulating his interest in quantum scattering. This research was supported by a Grant 
from the GIF, the German-Israeli Foundation for Scientific Research and Development.



\begin{thebibliography}{0}
\bibitem{note0} $\Gamma_{\rm cl}$ is calculated numerically from the exponential decay
of the classical probability to stay inside the sample.
\bibitem{note} The behavior of the extreme large $\Gamma$ tails of ${\cal P}(\Gamma)$ 
is essentially determined by the coupling to the leads which is model dependent. Their 
relative number is proportional to $M/L^2 \sim L^{-1}$ and therefore they are 
statistically insignificant.
\end{thebibliography}
\end{document}